\pdfoutput=1  
\documentclass[]{article}  
\usepackage{url,float}
\usepackage{graphicx}
\usepackage{amsfonts}
\usepackage{amssymb}
\usepackage{latexsym}

\newcommand{\hide}[1]{}

\newcommand{\ABox}{
\raisebox{3pt}{\framebox[6pt]{\rule{6pt}{0pt}}}
}
\newenvironment{proof}{{\bf Proof:}}{\hfill\ABox}

\newtheorem{theorem}{{\bf Theorem}}

\newtheorem{lemma}[theorem]{Lemma}

\newcommand{\lemlab}[1]{\label{lemma:#1}}

\newcommand{\eqlab}[1]{\label{eq:#1}}

\newcommand{\figlab}[1]{\label{fig:#1}}
\newcommand{\seclab}[1]{\label{sec:#1}}

\newcommand{\lemref}[1]{\ref{lemma:#1}}

\newcommand{\secref}[1]{\ref{sec:#1}}
\newcommand{\eqref}[1]{\ref{eq:#1}}
\newcommand{\figref}[1]{\ref{fig:#1}}

{\makeatletter
 \gdef\xxxmark{%
   \expandafter\ifx\csname @mpargs\endcsname\relax 
     \expandafter\ifx\csname @captype\endcsname\relax 
       \marginpar{xxx}
     \else
       xxx 
     \fi
   \else
     xxx 
   \fi}
 \gdef\xxx{\@ifnextchar[\xxx@lab\xxx@nolab}
 \long\gdef\xxx@lab[#1]#2{{\bf [\xxxmark #2 ---{\sc #1}]}}
 \long\gdef\xxx@nolab#1{{\bf [\xxxmark #1]}}
 \gdef\turnoffxxx{\long\gdef\xxx@lab[##1]##2{}\long\gdef\xxx@nolab##1{}}%
}

\def\P{{\mathcal P}}

\def\a{{\alpha}}
\def\b{{\beta}}
\def\th{{\theta}}

\newcommand{\squeezelist}{\setlength{\itemsep}{0pt}}

\title{Unfolding Restricted Convex Caps
}

\author{%
Joseph O'Rourke%
    \thanks{Dept. Comput. Sci., Smith College, Northampton, MA
      01063, USA.
      \protect\url{orourke@cs.smith.edu}.}
}

\begin{document}
\maketitle

\begin{abstract}
This paper details an algorithm for unfolding a class of convex
polyhedra,
where each polyhedron in the class
consists of a convex cap over a rectangular base,
with several restrictions:
the cap's faces are quadrilaterals, with vertices over an underlying integer
lattice, and such that the cap convexity is ``radially monotone,''
a type of smoothness constraint.
Extensions of Cauchy's arm lemma are used in the proof of non-overlap.
\end{abstract}

\section{Introduction}
\seclab{intro}
Few classes of convex polyhedra are known to be \emph{edge-unfoldable}:
unfolded by cutting along edges of the polyhedron and flattening into
the plane to a single piece without overlap.
Among the classes known to be are: pyramids, prismoids, and ``domes.''
See~\cite[Chap.~22]{do-gfalop-07} for background on this problem.

The purpose of this informal note is to introduce another narrow class of polyhedra
for which an edge-unfolding algorithm can be provided.
We call this class \emph{radially monotone lattice quadrilateral convex caps}.
The long name reflects the several qualifications needed to guarantee
correctness,
qualifications that hopefully can be removed by subsequent research.
The upper surface of such a polyhedron $\P$ is a 
\emph{convex cap} $C$ in the sense that
$C$ projects parallel to the $z$-axis to its 
base $B$ in the $xy$-plane without overlap.
In other words, the intersection of the cap $C$ with a line 
parallel to $z$ through an interior
point of $B$ is a single point.
In other terminology, $C$ is a \emph{terrain}.
$C$ is composed entirely of quadrilaterals, each of which projects to a unit lattice
square in the $xy$-plane, and whose projections tile $B$.
The base $B$ is restricted to be a rectangle.
$\P$ then is the convex hull of $C \cup B$, which fills in the four sides
$S_{x^-}$,
$S_{x^+}$,
$S_{y^-}$,
$S_{y^+}$.
See Figure~\figref{unf_k16_r2}(a,b).
The ``radially monotone'' qualification restricts the ``sharpness'' of
the convexity of the cap in a way that is not easily explained until we develop
more notation.

A more general shape would be a \emph{lattice convex cap},
the convex hull of $B$ and a set of points $(x,y,z)$
over each integer lattice point $(x,y)$ in $B$.  Here the faces are in general
triangles rather than quadrilaterals, and $C$ is a \emph{convex height field}.
The restriction to quadrilaterals
narrows the class considerably, as we now show.

\paragraph{Quadrilateral Restriction.}

Let the base $B$ range
over the lattice points
$x=1,\ldots,n_x$ and $y=1,\ldots,n_y$.
If the points above the front and left of $B$ are specified,
$$
c_x(1) \; : \; \{ (x,1) \; : \; x=1,\ldots,n_x \}
$$
and
$$
c_y(1) \; : \; \{ (1,y) \; : \; y=1,\ldots,n_y \}
$$
then all the points above the remainder of $B$ are determined.
This can be seen as follows.
If three of the four corners of a lattice cell are given,
$$
(x,y,z_0),
(x+1,y,z_1),
(x,y+1,z_3)
$$
then the fourth corner, $(x+1,y+1,z_2)$, is determined by the
plane containing the first three corner points, whose height
can be computed as
\begin{equation}
z_2 = -z_0 + z_1 + z_3
\eqlab{z2}
\end{equation}
Thus,
$$
(1,1,z_0),
(2,1,z_1),
(1,2,z_3)
$$
determine $(2,2,z_2)$,
and this determination propagates similarly out over the entire $n_x \times n_y$ rectangle.
Thus, if $n_x=n_y=n$, the $n^2$
lattice points of $B$ are fixed by specifying just the $2n -1$ values along 
the front and left sides.
A consequence of
Lemma~\lemref{convex.cap} below is 
that a lattice quadrilateral cap $C$ is convex if and only if
the curves at its two boundaries
$c_x(1)$ and $c_y(1)$
(i.e., the upper boundaries of $S_{x^-}$ and $S_{y^-}$)
are convex.

\paragraph{Radially Monotone.}
The full definition of when a convex cap is radially monotone will be
deferred until Section~\secref{radial} below, where it is first employed.
Here we define when a planar convex curve is radially monotone,
and give two sufficient conditions for radial monotonicity of a convex cap.
Let $c$ be a convex curve, and let $p_0$ and $p_i$ be two vertices of the curve,
with $p_{i+1}$ the vertex following $p_i$.
Then $c$ is \emph{radially monotone} if
the angle $\angle p_0 p_i p_{i+1} \ge \pi/2$,
for every pair of vertices $p_0$ and $p_i$.
The condition may be interpreted as requiring that
the next segment $p_i p_{i+1}$ of the chain
does not penetrate the circle of radius $|p_0 p_i|$
centered on $p_0$.
So points on the chain increase their radial distance from $p_0$.
(See ahead to Figure~\figref{radially_monotone}.)
The two sufficient conditions are as follows:
\begin{enumerate}
\squeezelist
\item
If the two convex curves $c_x(1)$ and $c_y(1)$ are both radially monotone,
then the convex cap is radially monotone.
\item
If each of these convex curves fits in a semicircle connecting its two endpoints,
then each is radially monotone, and so~(1) applies.
\end{enumerate}
This latter condition justifies the notion that radial monotonicity
enforces a type of smoothness on the convexity of the cap.

\section{Lattice Quadrilateral Cap Properties}
We now derive several properties of lattice quadrilateral convex caps.

\begin{lemma}[Parallelograms]
Every quadrilateral over a lattice cell is a parallelogram.
\lemlab{para}
\end{lemma}
\begin{proof}
Let the four corners of the quadrilateral be $a$, $b$, $c$, $d$
in counterclockwise order,
let $\a$ and $\b$ be the angles at $a$ and $b$ respectively,
and let $\Pi$ be the plane containing the quadrilateral.
We argue that the two coplanar triangles
$\triangle abd$ and 
$\triangle bcd$ are congruent.
They share the diagonal of the quadrilateral, $bd$.
The edge $ab$ must be congruent to edge $dc$, because both are 
the intersection of $\Pi$ with a unit vertical strip,
parallel (say) to the $xz$-plane.
Therefore $|ab|=|dc|$.
Similarly, $bc$ must be congruent to $ad$, because
both are produced by $\Pi$ intersecting the a unit vertical
$yz$-strip.
Therefore $|bc|=|dc|$.
Therefore $\triangle abd$ and 
$\triangle bcd$ have all edges the same length, and
so are congruent, as claimed.
So the angles at $a$ and at $c$ are the same, $\a$.
Similarly, the angles at $b$ and at $d$ are the same, $\b$.
\end{proof}

Call the chain of edges of $C$ above $(y,1),(y,2),\ldots,(y,n_x)$
the $x$-chain $c_x(y)$, and similarly define the $y$-chain $c_y(x)$.

\begin{lemma}[Cap Convexity]
For a lattice quadrilateral cap $C$,
all the $x$-chains are congruent to one another, as are all the $y$-chains.
\lemlab{convex.cap}
\end{lemma}
\begin{proof}
Consider two consecutive edges of the $x$-chain $c_x(y)$,
over $(x,y),(x+1,y),(x+2,y)$.
By the Parallelogram Lemma~(\lemref{para}),
the corresponding edges in the next $x$-chain $c_x(y+1)$ are
parallel to those in $c_x(y)$.
Thus the convex angle at the middle point $(x+1,y)$
is exactly the same as that at $(x+1,y+1)$.
Because this is true at every lattice point of each $x$-chain,
all are congruent.
\end{proof}

\noindent
The congruency of the chains is evident in Figure~\figref{unf_k16_r2}(a,b).

Let $z(x,y)$ be the point of $C$ above $(x,y)$.
\begin{lemma}[Cap Maxima]
The maximum $z$-height of
a lattice quadrilateral cap $C$ for
the points at a particular $x$-value,
$\max_y z(x,y)$,
is achieved at a $y$-value $y_{max}$ independent of $x$.
And the same claim holds when the roles of $x$ and $y$ are interchanged.
\lemlab{convex.cap.max}
\end{lemma}
\begin{proof}
This follows immediately from the congruency of the $x$-chains
provided by Lemma~\lemref{convex.cap}.
\end{proof}

\noindent
For example, in Figure~\figref{unf_k16_r2}(a,b), $y_{max}=14$.

\begin{figure}[htbp]
\centering
\includegraphics[width=\linewidth]{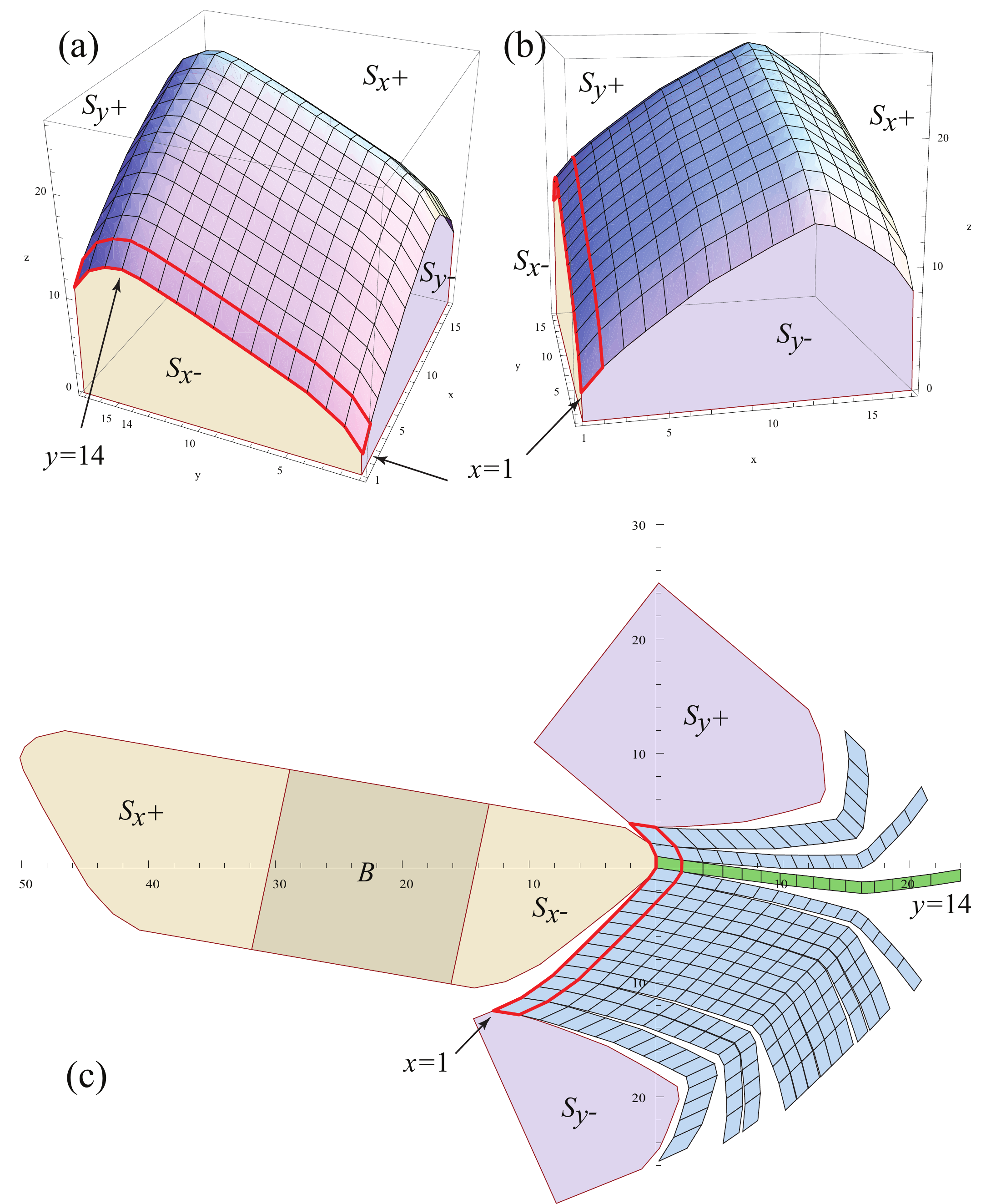}
\caption{(a,b) Two views of the same convex cap $C$.
The base $B$ is the square at $z=0$.
(c)~Unfolding of $\P$.}
\figlab{unf_k16_r2}
\end{figure}

\section{Sketch of Algorithm}
We first sketch the unfolding algorithm, using
Figure~\figref{unf_k16_r2} for illustration,
before justifying it in the following sections.
The strip of quadrilaterals between 
$y_{max}$ and $y_{max}+1$, call it
the \emph{$y_{max}$-strip} is unfolded with $(1,y_{max})$
at the origin and its leftmost edge
along the $y$-axis.  Thus it unfolds roughly horizontally.
See the green $y=14$ strip in Figure~\figref{unf_k16_r2}(c).
All the other ``parallel'' $x$-strips for different $y$-values
are arranged above and below this $y_{max}$-strip,
connected along the uncut edge between $x=1$ and $x=2$.
All other $x$-edges for $x \ge 2$ are cut.
We will show that these strips splay outward from the $y_{max}$-strip,
avoiding overlap.
The two side faces
$\{ S_{x^-}, S_{x^+},\}$
are unfolded attached to the base $B$,
and attached to the cap unfolding at the leftmost vertical
edge of the $y_{max}$-strip.
We will show that $S_{x^-}$ in particular does not overlap
with the $x=1,2$ strip (outlined in red in the figure).
The remaining two side faces,
$\{ S_{y^-},S_{y^+} \}$, are attached at the two ends of the $x=1$ strip
as shown in the figure.

\section{Quadrilateral Angles}

We will call the four angles of a quadrilateral
$(\a,\b,\a,\b)$ in counterclockwise order.
Note that $\a+\b=\pi$, so one angle determines all the others.
Much of our reasoning relies on an analysis of the behavior of
the quadrilateral angles.

Fix one quadrilateral over its lattice cell, with corner heights
$z_0=0,z_1,z_2,z_3$, counterclockwise
above $(0,0),(1,0),(1,1),(0,1)$ respectively,
as in Figure~\figref{alpha_z1}(a).
We will need detailed knowledge of how the $(0,0)$ corner
angle $\a$ varies as a function of $z_1$ and $z_3$:

\begin{lemma}[One Quad]
For $z_3 > 0$, $\a(z_1)$ is a strictly decreasing function
of $z_1$, passing through $\pi/2$ at $z_1=0$.
$\a(z_1,-z_3) = \pi/2 - \a(z_1,z_3)$,
so, for $z_3 < 0$, $\a(z_1)$ is a strictly increasing function
of $z_1$.
\lemlab{alpha.z1}
\end{lemma}
\begin{proof}
Explicit calculation shows that
$$
\a = \a(z_1,z_3) =
\cos^{-1} \left( \frac {- z_1 z_3}
{
\sqrt{ 1 + z_1^2 }
\sqrt{ 1 + z_3^2 }
}
\right)
$$
The claimed properties follow from the properties of the inverse cosine function.
See Figure~\figref{alpha_z1}(b).
\end{proof}

\begin{figure}[htbp]
\centering
\includegraphics[width=0.95\linewidth]{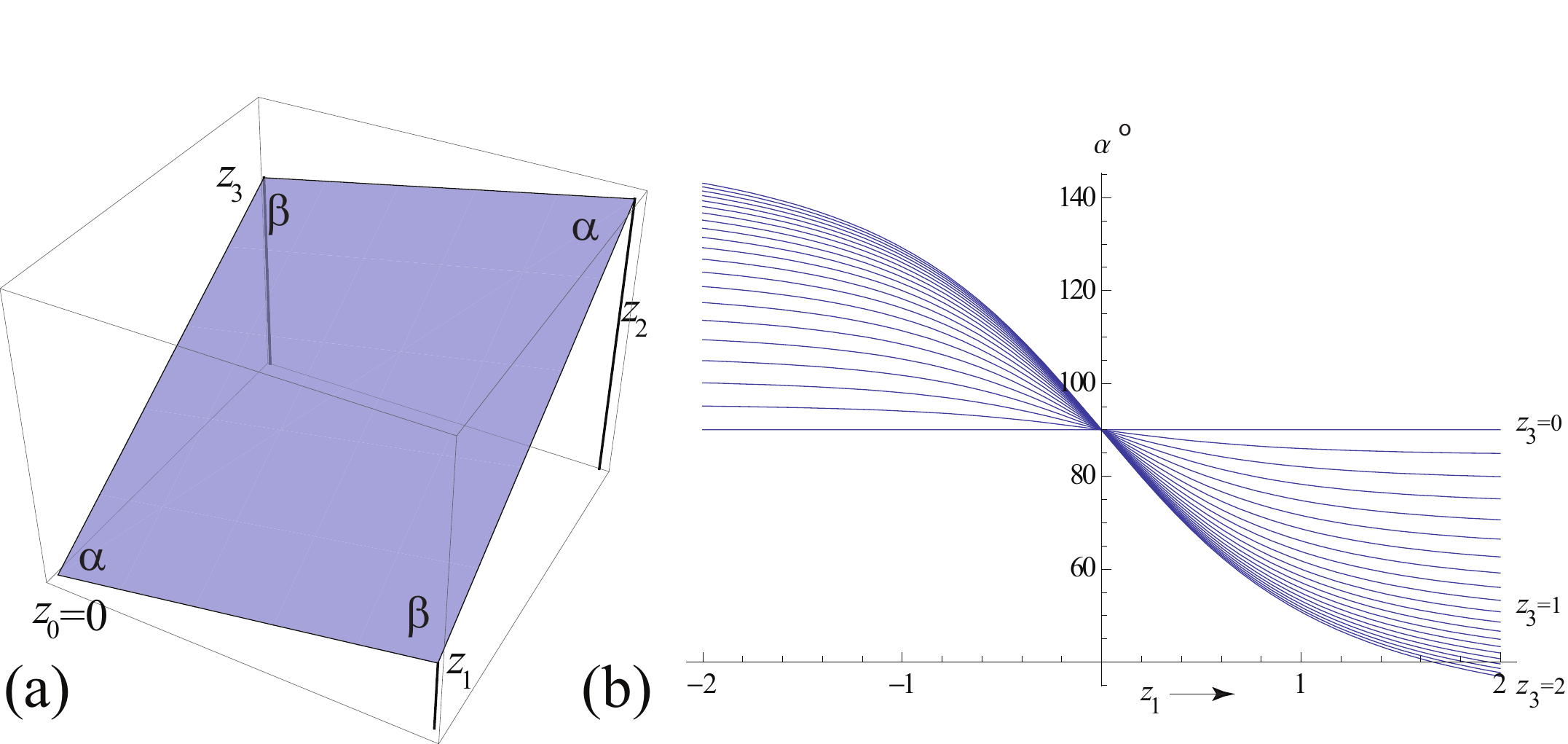}
\caption{(a) one quadrilateral;
(b)~$\a(z_1,z_3)$ is strictly decreasing for $z_3 > 0$.}
\figlab{alpha_z1}
\end{figure}

We now examine angles formed by two adjacent quadrilaterals.
We fix one in the position detailed above, and the other
adjacent to its left, with height $z'_1$ above the lattice point $(-1,0)$.
Note that the other three corners of this second quadrilateral are
then determined.
Let $\th$ be the angle in the vertical plane containing
the edges above $(-1,0),(0,0),(1,0)$,
and let $\a$ and $\b'$ be the angles of the quadrilateral incident
to the origin $(0,0,0)$.
See Figure~\figref{alphabeta_z3}(a).
We need the behavior of $(\a+\b')$ as $z_3$ varies:

\begin{lemma}[Two Quads]
Let the front edges of two adjacent quadrilaterals make
a convex angle of $\th$ in the vertical plane containing
those edges.
Then $(\a+\b')$ is a strictly increasing function of $z_3$,
passing through $\pi$ at $z_3=0$, and
with asymptote $\th$ for $z_3 < 0$ and $2\pi-\th$ for $z_3 > 0$.
\lemlab{alphabeta.z3}
\end{lemma}
\begin{proof}
A proof by explicit calculation, using
the formula used in the One-Quad Lemma~(\lemref{alpha.z1})
twice,
will not be presented.
Alternatively, one can arrive at the claim via the
One-Quad Lemma 
as follows.
From Figure~\figref{alpha_z1}, $\a$ decreases as $z_3$
increases, but, because
$|z'_1| > z_1$ by convexity, $\b'$ increases more than $\a$ decreases,
and so $\a+\b'$ increases.
See Figure~\figref{alphabeta_z3}(b) for sample plots.
Note that, as $z_3 \rightarrow +\infty$,
$\a$ and $\b'$ approach lying in the vertical plane above $\th$,
and so $\th + \a + \b' = 2 \pi$,
and as $z_3 \rightarrow -\infty$,
$\a + \b' \rightarrow \th$.
\end{proof}

\begin{figure}[htbp]
\centering
\includegraphics[width=0.95\linewidth]{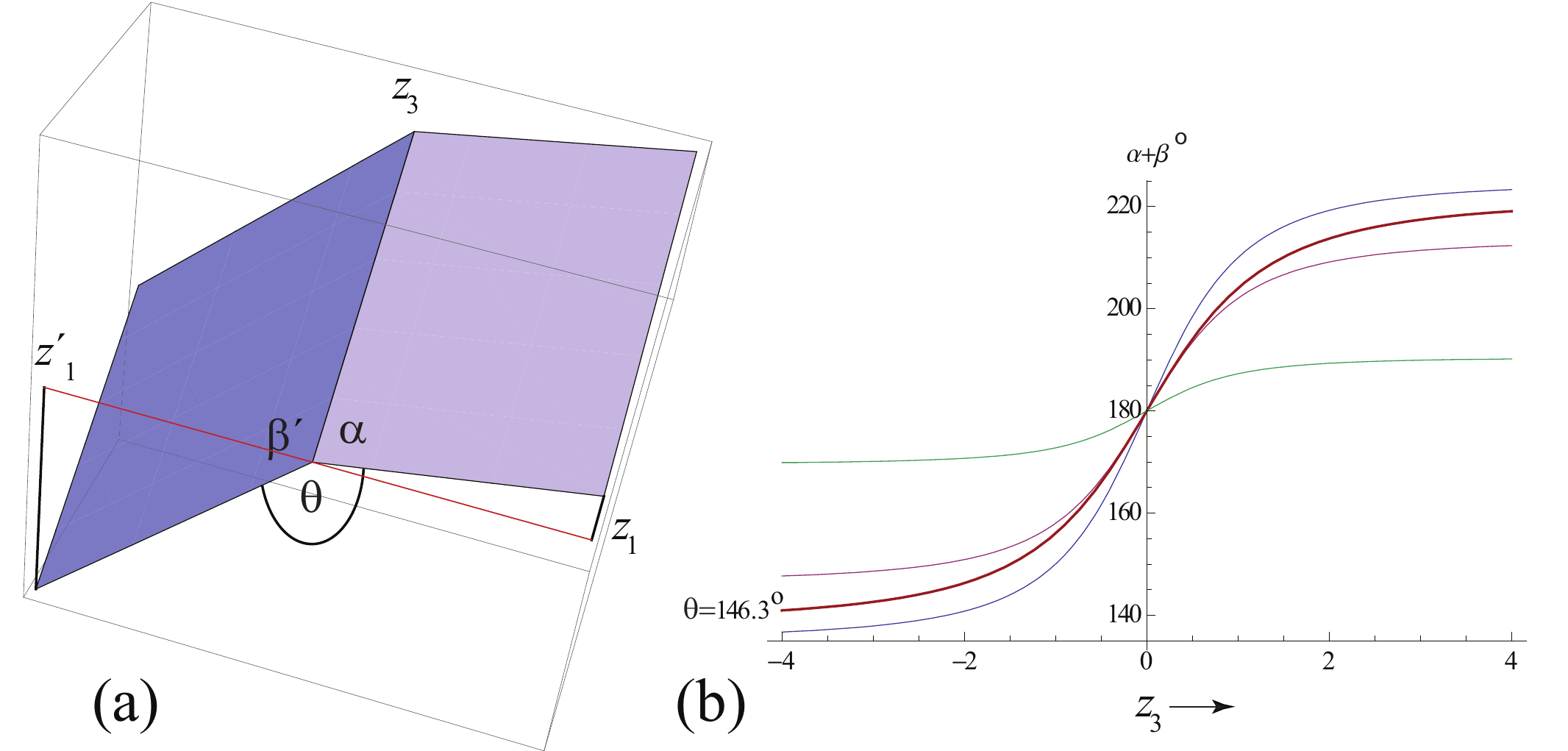}
\caption{(a) Two adjacent quadrilaterals;
here $\th=146.3^\circ$.
(b) $(\a+\b')$ is a strictly increasing function of $z_3$;
Note the asymptote $\th=146.3^\circ$ for the highlighted curve
corresponding to~(a).
}
\figlab{alphabeta_z3}
\end{figure}

Define the \emph{turn angle} $\tau$ at point of a curve with convex angle $\th$
to be $\pi - \th$.

\begin{lemma}[Strip Unfolding]
Each strip of quadrilaterals unfolds without self-overlap,
with convex boundary curves.
\lemlab{one.strip}
\end{lemma}
\begin{proof}
Let an $x$-strip lie between $y$ and $y+1$.
Let $\th(x)$ be the angle at $p=(x,y,z(x,y))$
in the vertical $y$-plane through $p$.
We know from the Convex Cap Lemma~(\lemref{convex.cap})
that $\th \le \pi$ because the curve is convex at $p$.
Let $\tau_0 = \pi - \th$ be the turn angle at $p$ in the vertical plane.
The turn angle of
the unfolded quadrilaterals incident to $p$
is $\tau = (\a+\b')-\pi$.
The Two-Quad Lemma~(\lemref{alphabeta.z3}) shows that $(\a+b') \in [ \th, 2\pi - \th ]$,
which implies that
$\tau \in [ \th - \pi, \pi - \th ]$,
i.e.,
$\tau \in [ -\tau_0, +\tau_0 ]$.
These are precisely the angle conditions for application of the
extension of Cauchy's arm lemma:
all turn angles of the convex curve are ``straightened''~\cite{o-ecala-01}.
(This is an extension of Cauchy's arm lemma because nonconvexities might be introduced.)
The usual conclusion is that, under such a \emph{valid reconfiguration},
the endpoints of the chain pull further apart.
In~\cite[Cor.~2]{o-dipp-03} this consequence is derived:
A valid reconfiguration of an open convex chain remains simple,
i.e., does not self-intersect.

Although in general a valid reconfiguration could result in a nonconvex
straightened curve,
in our case the Convex Cap Lemma~(\lemref{convex.cap}) 
places us  
to the left or right of $z_3=0$ in Figure~\figref{alphabeta_z3}(b),
keeping all turn angles $\tau$ the same sign, and therefore
resulting in a convex curve.

So now we have established that neither of the two unfolded boundary curves
of the $x$-strip,
call them $u_y$
from the chain at $y$ and $u_{y+1}$ that at $y+1$,
self-intersect.
But by the Parallelograms Lemma~(\lemref{para}),
$u_y$ and $u_{y+1}$ are connected by, and therefore separated by, edges that are all parallel.
(See Figures~\figref{unf_k16_r2}(c), \figref{16x16_unf}, and~\figref{32x32_unf}.)
Thus $u_y$ cannot intersect $u_{y+1}$.
Finally, these parallel ``ladder bars'' ensure that
it is not possible for the first $x$-segment to intersect
the last $x$-segment.
Therefore the entire strip unfolds without self-overlap.
\end{proof}

Finally we move to an analysis of four quadrilaterals, needed to compare
turn angles in the unfolding.

\begin{lemma}[Four Quads]
Let four quadrilaterals be incident to the origin $o=(0,0,0)$,
with those touching $y=1$ the \emph{upper quads}
and those touching $y=-1$ the \emph{lower quads}.
Let the upper quads unfold with turn angle $\tau$ at $o$,
and the lower quads with turn $\tau_0$ at $o$.
Then, if the middle edge of the lower quads, over $(0,-1),(0,0)$,
is uphill w.r.t. $z$ ($z'_3 < 0$ in the notation below), 
then $\tau_0 \ge \tau$:  the lower quads unfold to turn more sharply.
\lemlab{four.quads}
\end{lemma}

\begin{proof}
We extend the analysis in the Two-Quad Lemma~(\lemref{alphabeta.z3})
to four quadrilaterals incident to a central point,
taken to be $o=(0,0,0)$ without loss of generality.
We repeat all the notation developed in that lemma
and displayed in 
Figure~\figref{alphabeta_z3}(a),
and add two new quadrilaterals,
determined by one new parameter $z'_3 < 0$,
as shown in Figure~\figref{fourquads},
with angle labels as indicated.
Knowing from the Convex Cap Lemma~(\lemref{convex.cap}) that the
two convex $x$-chains are congruent, we know they determine the same
$\th$ in vertical planes, and so Lemma~\lemref{alphabeta.z3}
can be applied on an equal footing to both chains.
Our goal is to compare the angle $A = \a+\b'$
to $B=\a'_0 + \b_0$, for these angles determine the turn angles.
Note that $A$ is the same angle studied
in the Two-Quads Lemma~(\lemref{alphabeta.z3}), but $B = 2\pi -(\a_0 + \b'_0)$,
where $(\a_0 + \b'_0)$ is the angle to which Lemma~\lemref{alphabeta.z3} applies.

Now, in order to maintain convexity along the $y$-chain through the central
point, we must have $|z'_3| \ge z_3$
(recall the central point is the origin, so $z'_3 < 0$).
Because $|z'_3|$ is right of $z_3$ in 
Figure~\figref{alphabeta_z3}(b), and the angle sum is strictly increasing,
we know that $\a_0 + \b'_0 \ge \a+\b'$,
i.e., $2\pi-B \ge A$.
Let $\tau$ be the turn angle at $o$ of the unfolding
of the upper two quads, $\tau = A-\pi$,
and $\tau_0$ the same turn angle of the lower two quads, $\tau_0 = \pi-B$.
Substitution into the inequality $2\pi-B \ge A$
yields $\tau_0 \ge \tau$.
\end{proof}

\begin{figure}[htbp]
\centering
\includegraphics[width=0.45\linewidth]{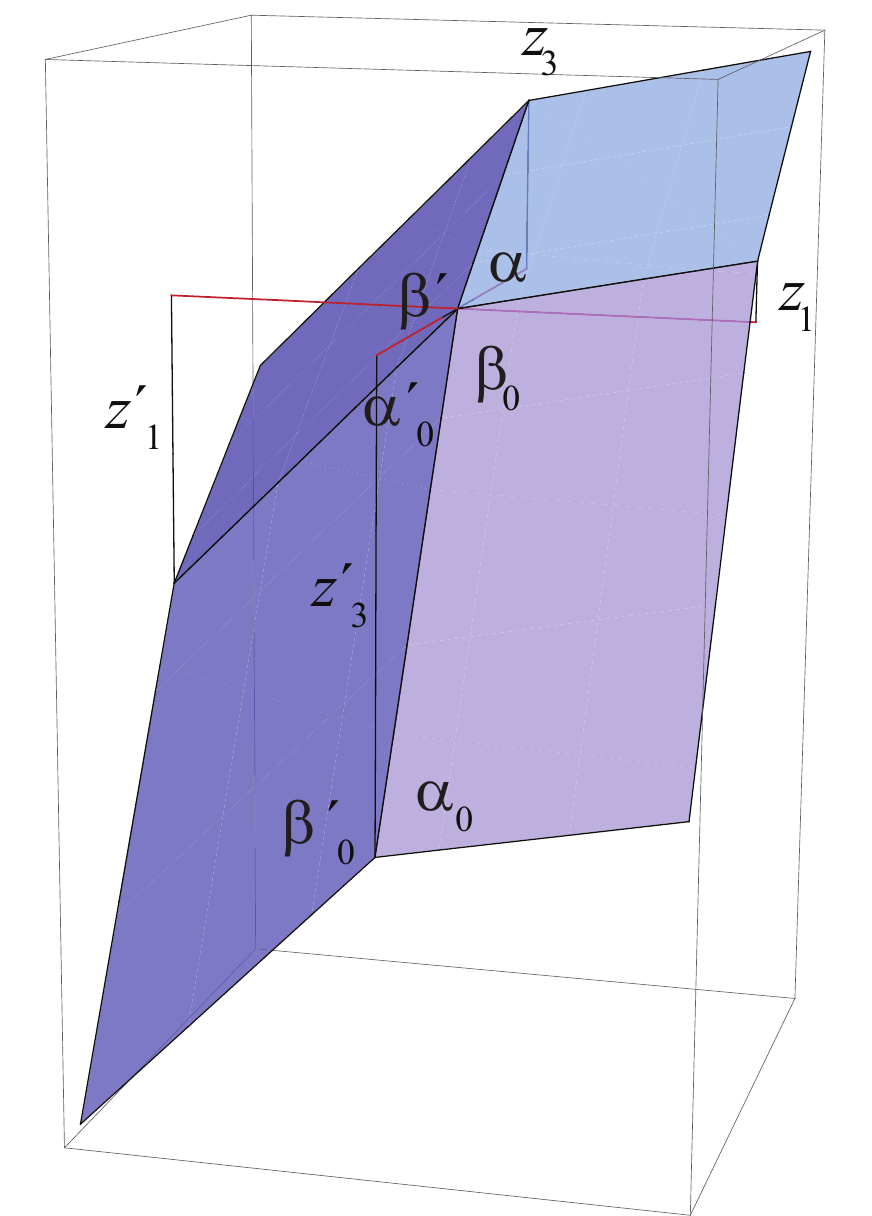}
\caption{Four quadrilaterals incident to $(0,0,0)$.
Convexity implies that $|z'_3| \ge z_3$.}
\figlab{fourquads}
\end{figure}

\noindent
The sharper turn angles guaranteed by this lemma provide a type
of separation between adjacent strips, but not enough to ensure
non-overlap.

\section{Radial Monotonicity}
\seclab{radial}
Now we finally come to the point where we need to use radial monotonicty,
to guarantee that the unfolding of adjacent strips do not overlap.
The condition we need is that the unfolded edge of each quadrilateral
strip be radially monotone, in the sense defined in Section~\secref{intro} above.
The Strip Unfolding Lemma~(\lemref{one.strip}) 
shows that the unfolded strip boundary is
a straightening of the curve $c_x(y)$ in the vertical plane,
so that radial monotonicity of $c_x(y)$ implies radial monotonicity of
the strip boundary, which justifies the first sufficient condition
claimed in Section~\secref{intro}.

We now show that any opening/straightening of a radially monotone curve
avoids overlap:\footnote{
   It may be that the equivalent of this lemma is available in the literature,
   but no explicit reference could be found.}

\begin{lemma}[Radial Monotonicity]
Let $c'$ be a straightening of a convex curve $c$, i.e.,
a reconfiguration such that every convex angle either stays
the same or opens closer to (but not exceeding) $\pi$.
Let $c$ lie in the positive quadrant with origin $o$,
with the left endpoints of both
$c$ and $c'$ at $o$.
If $c$ is radially monotone, then $c'$ does not intersect $c$ except at $o$.
\lemlab{radial}
\end{lemma}
\begin{proof}
First consider the curve $c_2$ in 
Figure~\figref{radially_monotone}, which is not
radially monotone at $a$ because $\angle oab < \pi/2$.
One can see that segment $ab$ cuts into the circle of radius
$|oa|$, and that under rotation about $o$, $a'b'$ crosses $ab$.
On the other hand, curve $c_1$ in the figure is radially monotone,
with each succeeding segment connecting two concentric circles.
Now rotation about $o$ moves each segment of $c_1$ on those circles,
and thus keeps the segment confined to the annulus between.  Therefore
such rotation cannot intersect any part of $c_1$.

Now because the lemma assumes that $c$ lies in the positive quadrant,
it is radially monotone from any other vertex $p$ beyond $o$, for the
angle $\angle pab$ can only be larger than $\angle oab$.
Now we view the opening of $c$ as composed of
rotations about each joint playing
the role of $o$ successively.  As the chain is rigid beyond
each rotation pivot $o$, the radial monotonicity
beyond $o$ is not affected.  So none of the rotations can cause
overlap, and the lemma is established.
\end{proof}
\begin{figure}[htbp]
\centering
\includegraphics[width=0.7\linewidth]{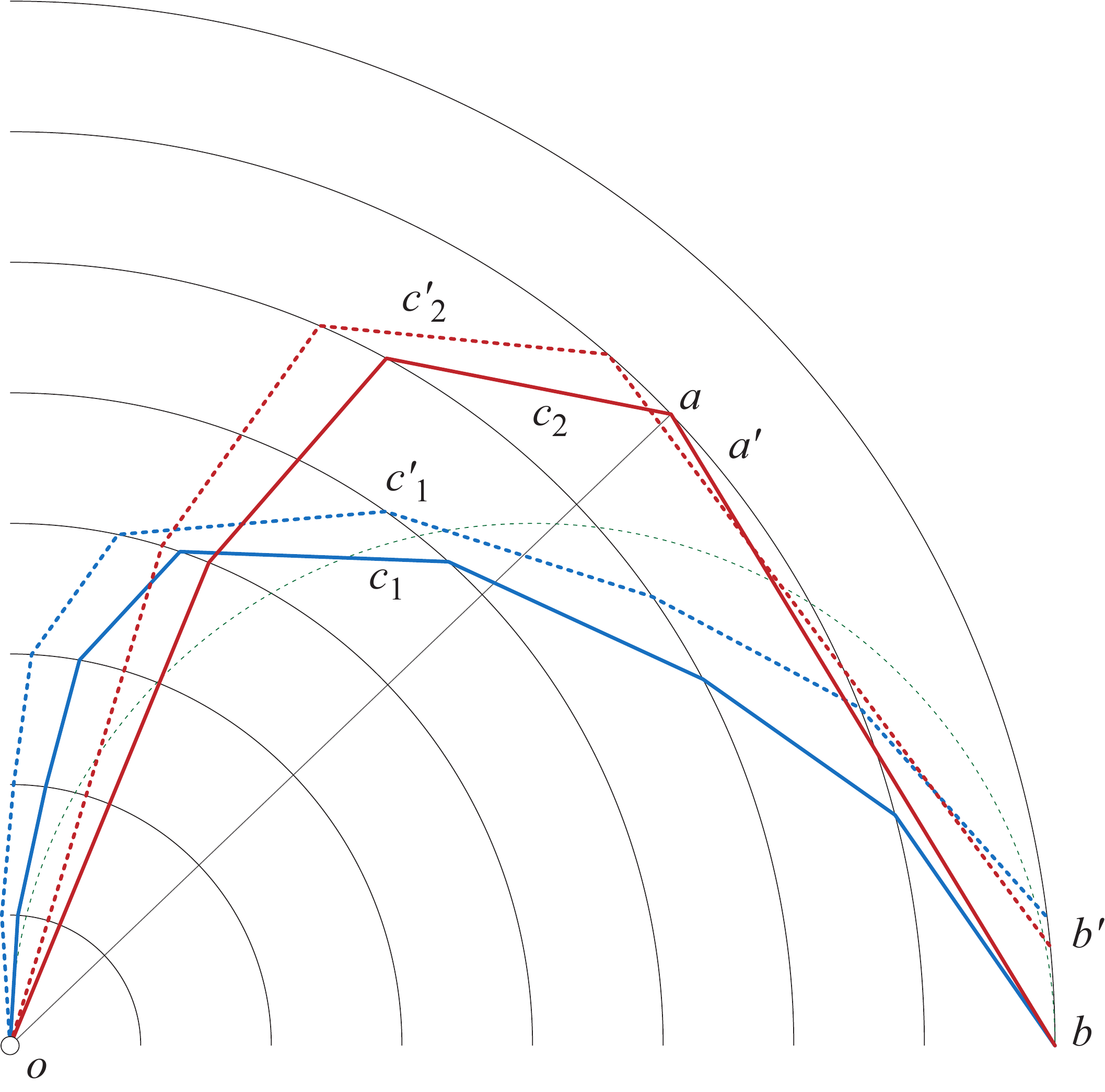}
\caption{$c_1$ is radially monotone, and $c'_1 \cap c_1 = \{o\}$; 
$c_2$ is not, and $ab \cap a'b' \neq \emptyset$.}
\figlab{radially_monotone}
\end{figure}

Note that if $c$ stays inside the semicircle 
connecting its endpoints (drawn lightly in Figure~\figref{radially_monotone}),
then radial monotonicity is guaranteed (because every point on the semicircle
subtends $\pi/2$ from the diameter).
This justifies the second sufficient condition
claimed in Section~\secref{intro}.
(None of the various sufficient conditions mentioned are necessary,
e.g., $c_1$ in the figure does not fit in the semicircle but is nevertheless
radially monotone.)

\begin{lemma}[Adjacent Strips]
Adjacent $x$-strips of quadrilaterals unfold without overlap.
\lemlab{adj.strips}
\end{lemma}
\begin{proof}
This now follows immediately from 
the Four-Quad Lemma~(\lemref{four.quads}),
which establishes that adjacent strip boundaries correspond
to a convex curve $c$ and a straightening $c'$
(because $\tau_0$, corresponding to $c'$, is $\ge \tau$,
corresponding to $c$),
and 
the Radial Monotonicity Lemma~(\lemref{radial}), 
which guarantees non-overlap of the two curves.
\end{proof}

\noindent
\begin{sloppypar}
Clearly the same reasoning guarantees the non-overlap of
the side faces $S_{x^-},S_{y^-},S_{y^+}$ with their adjacent
quadrilateral strips.
See Figures~\figref{16x16_unf} and~\figref{32x32_unf} for further
examples of cap unfoldings.
\end{sloppypar}

That the radial monotonicty restriction is sometimes necessary is
established by the overlapping unfolding of a convex cap that has such a
plummet in $z$ values near one corner of $B$ that the cap curves fail to be
radially convex there.
See Figure~\figref{overlap_r2}.
\begin{figure}[htbp]
\centering
\includegraphics[width=0.9\linewidth]{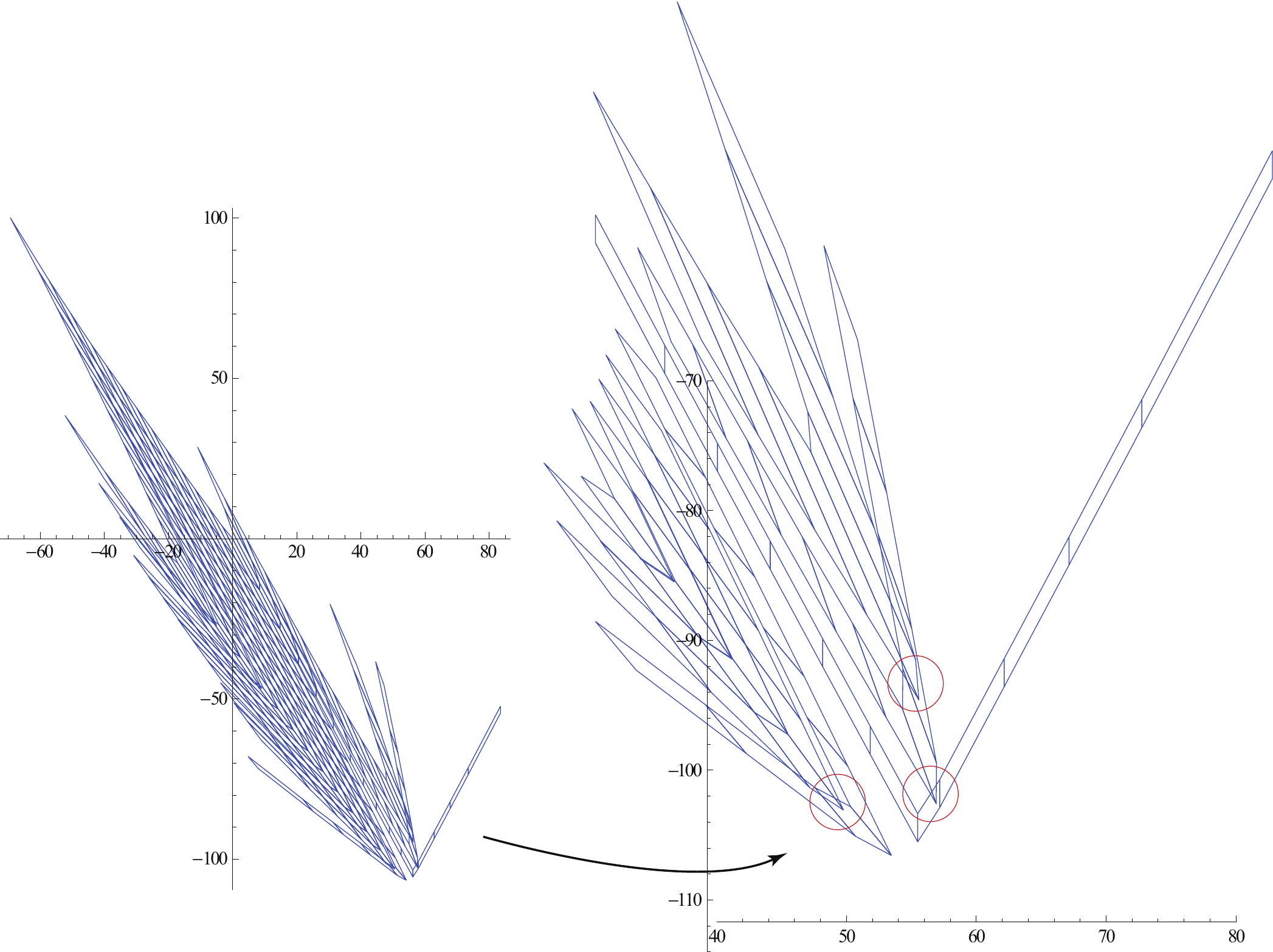}
\caption{Overlap caused by a sharply convex cap.
(Side faces are not shown.)
}
\figlab{overlap_r2}
\end{figure}

\section{Global Non-Overlap}
We have now established that all overlap is avoided locally, but there
still remains the issue of possible ``global overlap,'' overlap of widely separated
portions of the surface.
It is easy to add rays to the ends of each $x$-strip, and
rays on the sides of $S_{y^-}$ and $S_{y^+}$, that shoot to
infinity without crossing, as in
Figure~\figref{unf_rays}.
The $x$-strip rays can be thought of as edges of extensions of $c_x(y)$
with long ``tail'' quadrilaterals extending toward $z \rightarrow -\infty$,
which can be seen to preserve radial monotonicity.
These rays partition the plane and separate the pieces of the unfolding from
one another.
\begin{figure}[htbp]
\centering
\includegraphics[width=0.5\linewidth]{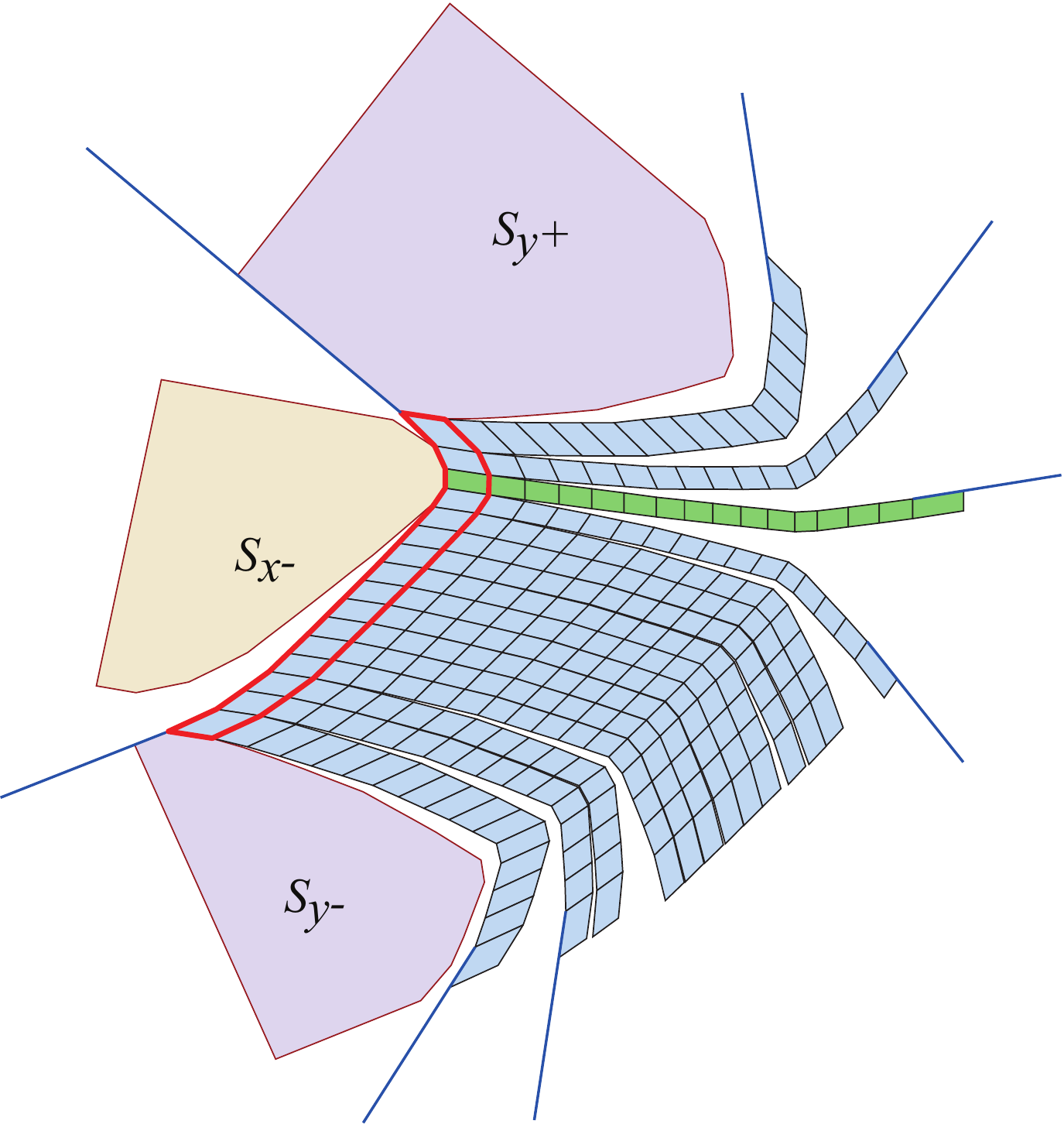}
\caption{Extension by rays establishes global non-overlap.}
\figlab{unf_rays}
\end{figure}

\section{Future Work}
The class of shapes for which the presented unfolding algorithm
guarantees non-overlap is narrow and of no interest except as
a possible stepping stone to more general shapes.
Next steps
(in order of perceived difficulty)
include attempting to
extend the algorithm to:
\begin{enumerate}
\squeezelist
\item remove the radially monontone assumption;
\item remove the assumption that $B$ is a rectangle;
\item include shapes with both an upper and a lower convex cap surface;
\item remove the assumption that all faces are quadrilaterals.
\end{enumerate}

\begin{figure}[htbp]
\begin{minipage}[b]{0.48\linewidth}
\centering
\includegraphics[width=\linewidth]{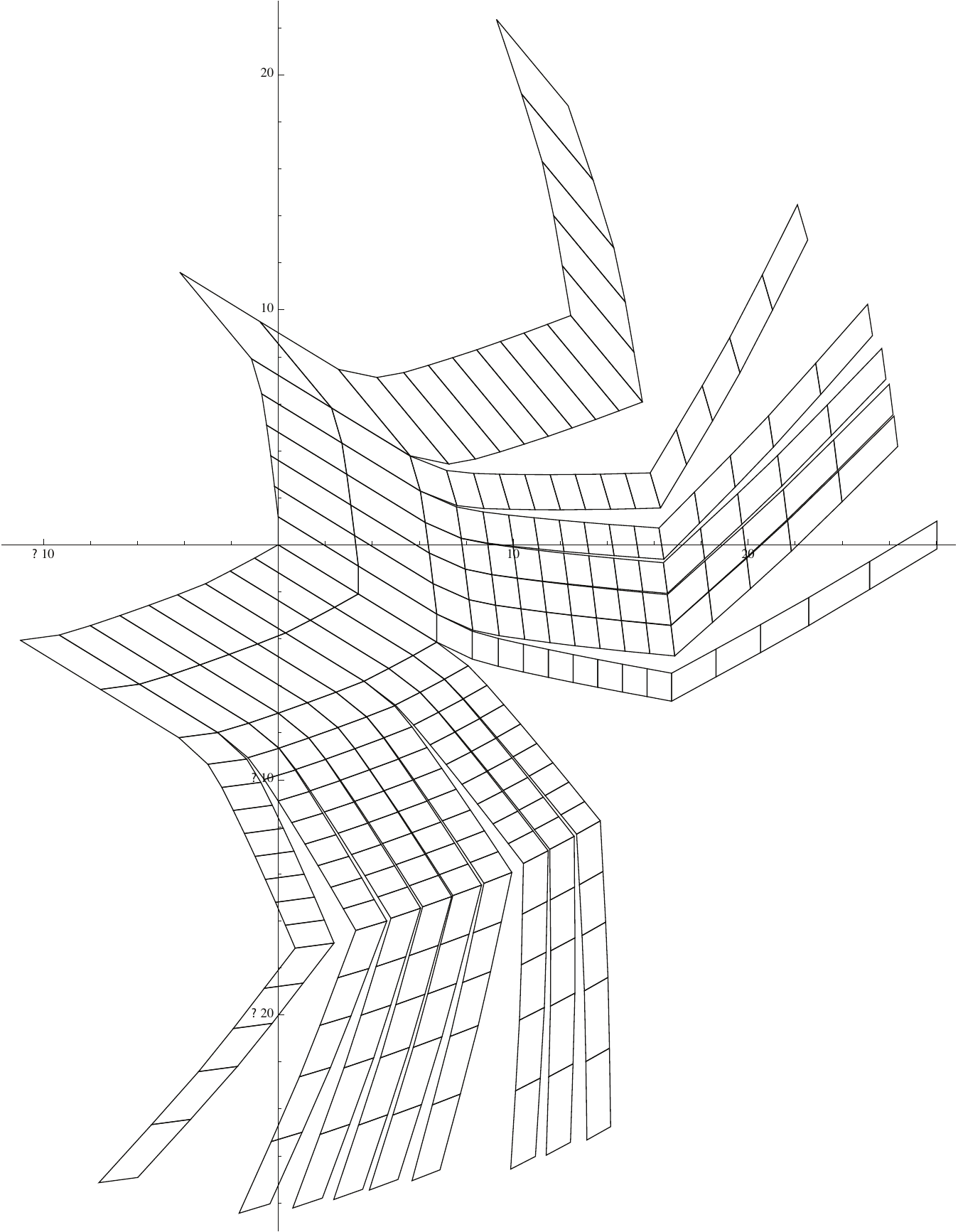}
\caption{$16 \times 16$ cap unfolding.}
\figlab{16x16_unf}
\end{minipage}%
\hspace{0.04\linewidth}%
\begin{minipage}[b]{0.48\linewidth}
\centering
\includegraphics[width=\linewidth]{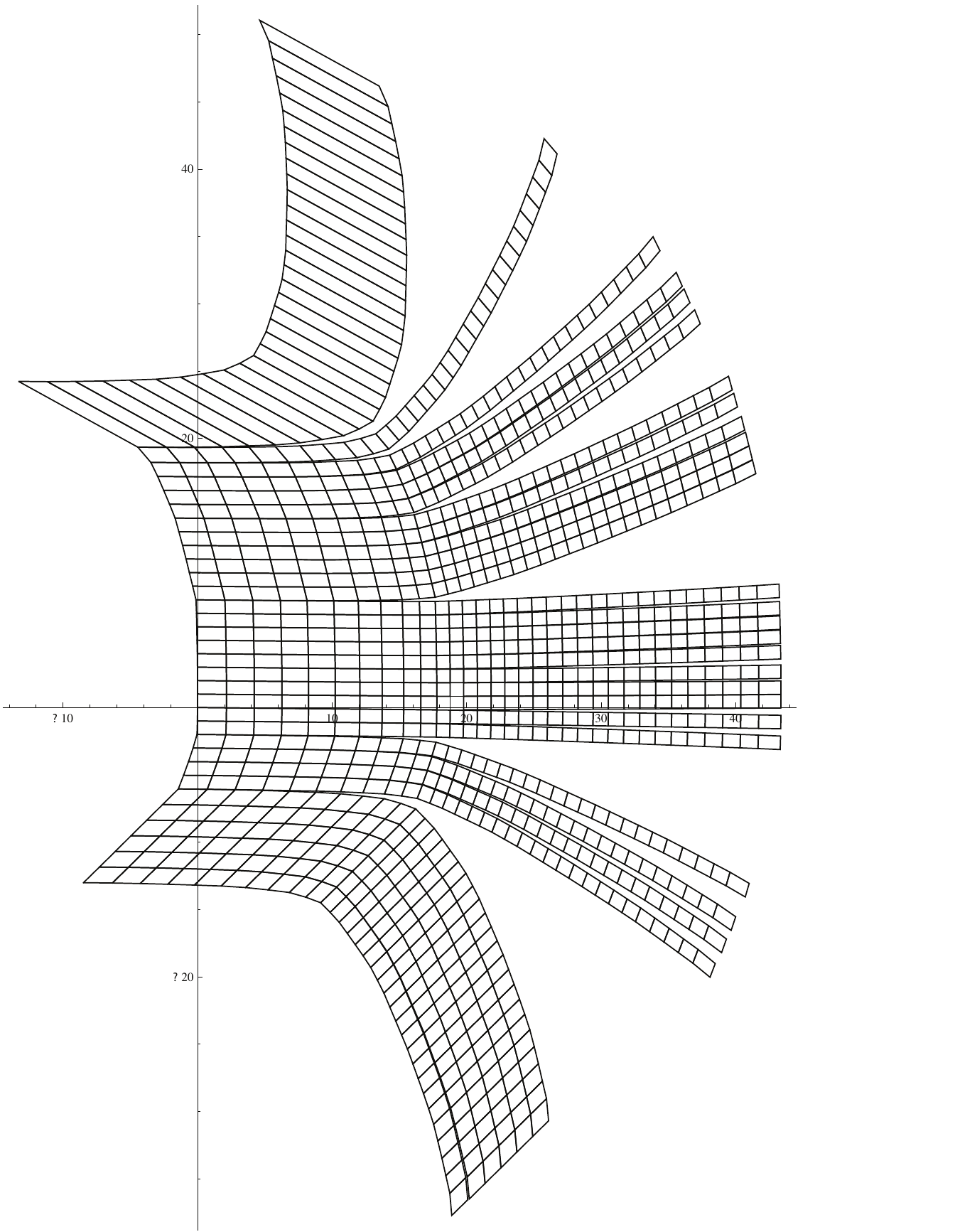}
\caption{$32 \times 32$ cap unfolding.}
\figlab{32x32_unf}
\end{minipage}%
\end{figure}


\bibliographystyle{alpha}
\bibliography{/home/orourke/bib/geom/geom}

\begin{thebibliography}{O'R03}

\bibitem[DO07]{do-gfalop-07}
Erik~D. Demaine and Joseph O'Rourke.
\newblock {\em Geometric Folding Algorithms: Linkages, Origami, Polyhedra}.
\newblock Cambridge University Press, July 2007.
\newblock \url{http://www.gfalop.org}.

\bibitem[O'R01]{o-ecala-01}
Joseph O'Rourke.
\newblock An extension of {Cauchy}'s arm lemma with application to curve
  development.
\newblock In {\em Proc. 2000 Japan Conf. Discrete Comput. Geom.}, volume 2098
  of {\em Lecture Notes Comput. Sci.}, pages 280--291. Springer-Verlag, 2001.

\bibitem[O'R03]{o-dipp-03}
Joseph O'Rourke.
\newblock On the development of the intersection of a plane with a polytope.
\newblock {\em Comput. Geom. Theory Appl.}, 24(1):3--10, 2003.

\end{thebibliography}
\end{document}